\documentclass[12pt]{article}
\usepackage{graphicx}
\usepackage{amsmath,amsfonts,amssymb}
\usepackage{bbm}
\usepackage[round]{natbib}
\usepackage{comment}
\usepackage{subcaption}
\usepackage{xr}
\usepackage{authblk}

\usepackage{setspace}

\begin{document}

\title{Composite Dyadic Models for Spatio-Temporal Data}
\date{June 3, 2024}

\begin{comment}
\author[1]{Michael R. Schwob}
\author[1]{Mevin B. Hooten}
\author[1,2,3]{Vagheesh Narasimhan}
\affil[1]{Department of Statistics and Data Sciences, \authorcr The University of Texas at Austin}
\affil[2]{Department of Integrative Biology, The University of Texas at Austin}
\affil[3]{Department of Population Health, Dell Medical School, \authorcr The University of Texas at Austin}
\end{comment}

\author{Michael R. Schwob$^{1,*}$, Mevin B. Hooten$^{1}$, and Vagheesh Narasimhan$^{1,2,3}$\\
$^{1}$Department of Statistics and Data Sciences, UT Austin\\
$^{2}$Department of Integrative Biology, UT Austin\\
$^{3}$Department of Population Health, Dell Medical School\\
$^{*}$\textit{email:} schwob@utexas.edu}

\doublespacing

\maketitle
\begin{abstract}
    Mechanistic statistical models are commonly used to study the flow of biological processes. For example, in landscape genetics, the aim is to infer spatial mechanisms that govern gene flow in populations. 
    Existing statistical approaches in landscape genetics do not account for temporal dependence in the data and may be computationally prohibitive.
    We infer mechanisms with a Bayesian hierarchical dyadic model that scales well with large data sets and that accounts for spatial and temporal dependence. 
    We construct a fully-connected network comprising spatio-temporal data for the dyadic model and use normalized composite likelihoods to account for the dependence structure in space and time.
    We develop a dyadic model to account for physical mechanisms commonly found in physical-statistical models and apply our methods to ancient human DNA data to infer the mechanisms that affected human movement in Bronze Age Europe.
\end{abstract}

\section{Introduction}

%% new introduction
% starts with applications to motivate need for a solid method
Mechanistic statistical models are designed to capture the underlying processes that generate observed data and have been used in atmospheric sciences \citep{wikle2001spatiotemporal}, ecology \citep{lu2020nonlinear}, landscape genetics \citep{hanks2013circuit}, and spatial epidemiology \citep{hefley2017dynamic}. For example, spatial epidemiologists use mechanistic models to infer the effect of environments on disease spread. Similarly, the field of landscape genetics is focused on understanding how landscape connectivity affects gene flow in populations of organisms. In these applications, the focus is on inferring the spatio-temporal process of interest and, consequently, the structure of the observed data in space and time. Physical-statistical models are commonly used to study such processes by embedding (partial) differential equations at the process level of a hierarchical statistical framework \citep{berliner2003physical,berliner2023excursions}. We present a framework that is motivated by differential equations, but that assumes a phenomenological form commonly used to study the flow of processes: dyadic regression. 

Dyadic regression is often presented in the context of network analysis. Given a spatio-temporal data set, a network may be constructed with nodes comprising the spatially- and temporally-referenced observations. An edge connects each of these observations to form a fully-connected network $\mathcal{N}=(\mathcal{V},\mathcal{E})$, where $\mathcal{V}$ and $\mathcal{E}$ denote the set of nodes and edges, respectively. 
The field of network analysis has experienced an explosion in scalable methodologies to meet the rapid growth in network sizes \citep{kolaczyk2014statistical}. 
One such methodology is dyadic regression, where pairwise outcomes of interest (i.e., edge weights) are regressed on node-level characteristics. In general, dyadic models are used to analyze data comprised of pairs of nodes (i.e., dyads), and they are often used to study relationships, interactions, and connections between pairs of nodes \citep{kenny2020dyadic}. Dyadic regression allows us to analyze how node-level covariates can predict pairwise outcomes of interest, and spatial dyadic regression has been used in many fields to regress pairwise outcomes on spatial features observed at the node-level, including landscape genomics \citep{wang2013examining}, international relations \citep{graham2020dyadic}, and infectious disease transmissions \citep{warren2023spatial}. 

We extend conventional dyadic regression models by motivating them with an underlying physical process model (a stochastic differential equation) that motivates the flow of the process in any parametric space. 
The connection between stochastic differential equations (SDEs) and dyadic regression implies that conventional dyadic regression models may be used to infer the flow of processes that gave rise to the data; this significantly improves their interpretability and motivates their adoption in studies concerning the flow of processes in any parametric space. We demonstrate our proposed dyadic model and leverage its connection with SDEs to study landscape genetics. In particular, we seek to infer the spatio-temporal mechanism that governed dispersal and migration in human populations in Bronze Age Europe, where the process of interest is the movement of genes and individuals.

% current landscape genetics model
A common approach to landscape genetics involves the use of circuit theory to infer functional connectivity, where spatial data are treated as a spatially-referenced circuit \citep{mcrae2008using}. The circuit can be viewed as a network comprising resistors as nodes and pairwise resistance distances as the effective distances between nodes, where resistance distance is usually defined as a function of characteristics in the parametric space. 
Statistical circuit theoretic approaches are formally coupled with spatial dynamics and are computationally feasible under standard Markov assumptions \citep{hanks2013circuit}.
The proposed framework shares these traits and also allows us to account for temporal proximity to isolate processes of interest. Additionally, the proposed method scales well for large data sets even when deviating from standard Markov assumptions. 

However, a consequence in constructing a fully-connected network for the dyadic regression is that we form a data set that may contain edges that are not meaningful to study the process of interest. Tobler's first law of geography assumes that processes are decreasingly related as their distance increases \citep{miller2004tobler}. Similarly, as the temporal lag between nodes increases, the nodes may be decreasingly related. Common procedures following the construction of a fully-connected network are subnetwork identification or graph reduction, where insignificant edges are eliminated from the network \citep{nguyen2019comprehensive}; formal tests for nodal dependence can aid such pursuits \citep{fosdick2015testing}. Alternatively, the graphical lasso can be used to estimate sparse precision matrices and identify subgraphs \citep{friedman2008sparse}. We let the model reduce the influence of less relevant edges on posterior inference by weighting the edge-level data using heterogeneous composite likelihoods. Composite likelihoods (also called fractional or power likelihoods) assign composite weights to the data model, resulting in weighted posterior inference \citep{holmes2017assigning,miller2018robust}. We combine composite likelihoods with dyadic analysis and relate the proposed composite weighting scheme to standard regularization techniques. 
We also assess the use of composite likelihoods to improve posterior inference for latent spatio-temporal mechanisms. 

\section{Methods}

We denote a fully-connected network $\mathcal{N}=(\mathcal{V},\mathcal{E})$, where $\mathcal{V}=\{v_1,...,v_n\}$ is the set of $n$ nodes and $\mathcal{E}=\{e_{12},...,e_{(n-1)n}\}$ is the set of $N={n \choose 2}$ edges. Each node is assumed to be spatially- and temporally-referenced. We let $y_i$ and $\mathbf{x}_i\equiv\mathbf{x}(\mathbf{s}_i)$ denote the response and covariates at node $i$, respectively, where $\mathbf{s}_i$ denotes the location of node $i$.
We let $\widetilde{y}_{ij}$ denote the weight of edge $e_{ij}$, which connects nodes $i$ and $j$; $\widetilde{y}_{ij}$ may be observed directly or computed as a dissimilarity between the node-level responses $y_i$ and $y_j$. In dyadic regression, the edge weight $\widetilde{y}_{ij}$ is treated as the dyadic outcome between nodes $i$ and $j$. 

\subsection{Connecting SDEs with Dyadic Regression}

\begin{figure}
    \centering
    \includegraphics[width=\textwidth]{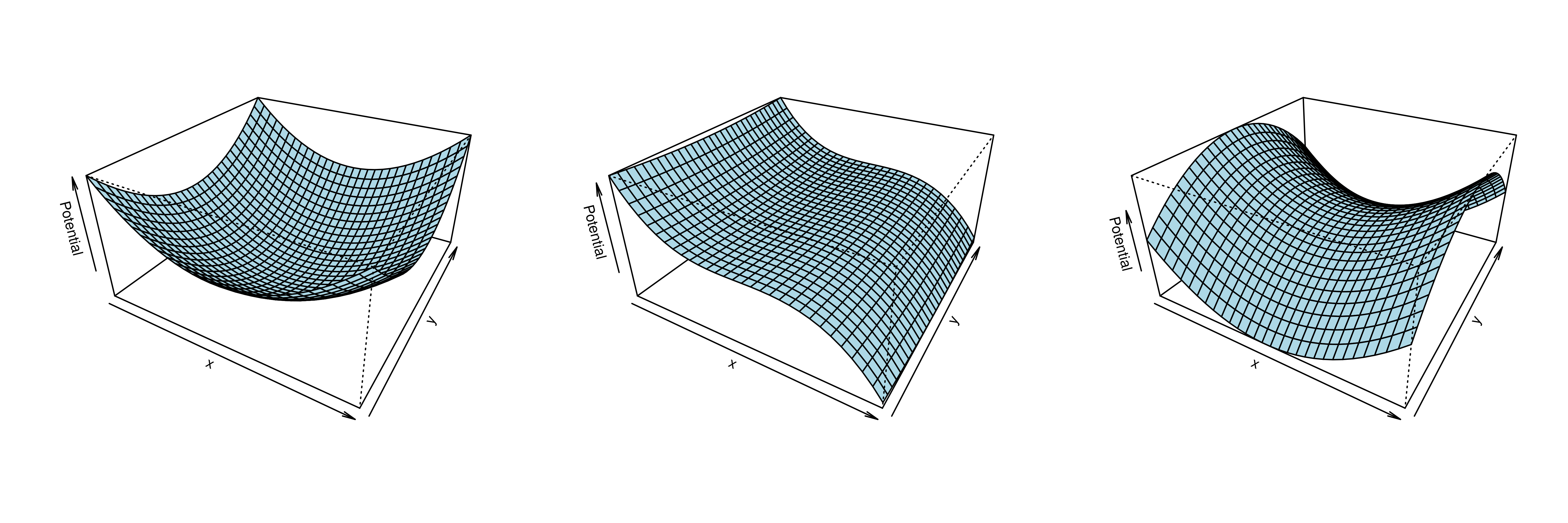}
    \caption{Examples of potential surfaces.}
    \label{fig:potPlot_examples}
\end{figure}

Mechanisms that give rise to the pattern of spatial processes are often visualized as potential surfaces, where the gradient of the surface dictates the resistance to the flow of the process \citep{hooten2017animal}. Regions with higher potential are more resistant to the flow of the process, and the process flows more freely in lower potential regions (see Figure \ref{fig:potPlot_examples}). 
Potential surfaces have been used to visualize spatial mechanisms in animal movement \citep{brillinger2012use}, economics \citep{puu1981structural}, and environmental sciences \citep{dunphy2006influence}. 
We aim to construct our Bayesian hierarchical model with an embedded potential surface that is a function of model parameters and processes.

%%%%%%%%% Updated (Simplified) Notation
Potential surfaces are often obtained via differential equations \citep{teller1937crossing}. We consider a stochastic differential equation written as a Langevin equation of the form
\begin{equation}\label{eq:basicSDE}
    \frac{d\mu(t)}{dt} = g(\mu(t)) + \frac{\epsilon(t)}{dt},
\end{equation}
where $\mu(t)$ is the process of interest, $g(\mu(t))$ is the advection component of the SDE, $\epsilon(t)/dt$ is the diffusion component of the SDE, and $\epsilon(t)$ is a noise process that is often assumed to be Gaussian. 
The SDE (\ref{eq:basicSDE}) is an advection-diffusion differential equation and is commonly used to study the flow of processes through space and time. The advection component $g(\mu(t))$ represents the flow of the process via the medium in which the process occurs. The diffusion component $\epsilon(t)/dt$ represents the spreading of the process and is usually a result of random motion.
In landscape genomics, the advection component represents the flow of genes (the process) due to the landscape (the medium), and the diffusion component represents the spread of genes unaffected by the landscape. Other applications of advection-diffusion differential equations include the movement of animals \citep{hooten2017animal} and the spread of pollutants in the atmosphere or water \citep{daniela2012air}. Although we consider a univariate process $\mu(t)$, the proposed methods may be extended for multivariate processes.

The SDE (\ref{eq:basicSDE}) is defined for a continuous process $\mu(t)$; however, many spatio-temporal data sets comprise discrete observations of the process: $\{\mu_t\}$, $t=1,\ldots,T$, where the subscript implies a discrete temporal indexing as opposed to the continuous indexing $\mu(t)$. 
We restrict our attention to discretely observed data sets and consider the Eulerization of (\ref{eq:basicSDE}): 
\[
    \mu_t - \mu_{t-dt} = g(\mu_{t-dt})dt + \epsilon_t,
\]
which relates the change of the process after a $dt$ time-step with advection and diffusion processes. 
Although the discretization of time assumes a constant rate of change within each time-step, the Eulerization of SDEs is common for analyzing discretely observed processes.
\cite{brillinger2012use} showed that a potential function $p(\mu_t)$ could be used to model advection such that $g(\mu_{t-dt}) =dp(\mu_t)/dt$.
Evaluating the potential function $p(\mu_t)$ at every point in the domain (i.e., every time) renders the potential surface that visualizes the flow of the process. 
We introduce the potential function to the discretized SDE to obtain the potential surface and write the advection-diffusion SDE (\ref{eq:basicSDE}) as the Euler discretization
\begin{equation}\label{eq:disc}
    \mu_t - \mu_{t-dt} = p(\mu_t) - p(\mu_{t-dt}) + \epsilon_t.
\end{equation}
The specification in (\ref{eq:disc}) relates the change in the process of interest to the change in the potential function; this intuition forms the foundation for our method to infer spatio-temporal processes. 
We note that, if the process of interest is observed continuously, the continuous analogue of (\ref{eq:disc}) may be used:  $d\mu(t)/dt=-\nabla p(\mu(t)) + \epsilon(t)/dt,$ where $\nabla$ is the gradient operator \citep{hooten2017animal}.

% relating the ADE to the dyadic network
We relate dyadic regression to the discretized advection-diffusion differential equation in (\ref{eq:disc}). First, we let $t_i$ denote the temporal reference of node $i$, and we define $\widetilde{y}_{ij}=y_j-y_i$ for all pairs of nodes, where $t_j>t_i$; when times are equivalent, an arbitrary order is assigned and preserved throughout the analysis. Conventional dyadic regression models relate the dyadic outcome $\widetilde{y}_{ij}$ to the difference in a function of node-level covariates: 
\begin{equation}\label{eq:analogue}
    \widetilde{y}_{ij} = \rho(\mathbf{x}_j) - \rho(\mathbf{x}_i) + \epsilon_{ij},
\end{equation}
where $\rho(\mathbf{x}_i)$ is a function of covariates at node $i$ (i.e., $\rho(\mathbf{x}_i)=\mathbf{x}_i'\boldsymbol{\beta}$) and $\epsilon_{ij}$ is Gaussian noise. 
We note that $\widetilde{y}_{ij}$ is the change in the process of interest and $\rho(\mathbf{x}_j) - \rho(\mathbf{x}_i)$ is the change in a function of node-level covariates between discrete observations. Thus, (\ref{eq:analogue}) is analogous to (\ref{eq:disc}), and the mean structure of conventional dyadic models can represent the gradient of a potential function, thereby describing a potential surface. In what follows, we extend (\ref{eq:analogue}) to account for spatio-temporal dependence between nodes and dyadic dependence between edges.

\subsection{Composite Likelihoods}

% discuss the importance of correlation for learning about the potential surface
Spatio-temporal statistical models assume that observations are decreasingly dependent as spatial and temporal lags increase \citep{cressie2015statistics}. Thus, dyads associated with nodes that are distant in space or time may inaccurately represent the flow of the process, hence obscuring the mechanism. We expect to learn more about the potential surface with dyads comprising correlated nodes than dyads comprising uncorrelated nodes. Therefore, we use composite likelihoods to attenuate the contribution of dyadic outcomes on posterior inference.

% introduction to composite likelihoods
Composite likelihoods have become popular in Bayesian statistics, where likelihoods are raised to a composite weight $w$ to allow for Bayesian learning under model misspecification \citep{holmes2017assigning}, to robustify Bayesian inference via coarsening \citep{miller2018robust}, or to account for overdispersion in count data \citep{fletcher2023simple}. 
For example, we consider the misspecified model $[\mathbf{y}\mid\boldsymbol{\alpha}]$ with prior $[\boldsymbol{\alpha}]$, where $[\cdot\mid\cdot]$ represents the conditional probability density or mass function \citep{gelfand1990sampling}. Then, a coherent and formal Bayesian update of the prior $[\boldsymbol{\alpha}]$ to the posterior $[\boldsymbol{\alpha}\mid\mathbf{y}]$ exists and is of the form
\begin{equation}\label{eq:pw}
    [\boldsymbol{\alpha}\mid\mathbf{y}]_w \propto [\mathbf{y}\mid\boldsymbol{\alpha}]^w[\boldsymbol{\alpha}],
\end{equation}
where $w$ is a composite weight that calibrates the two loss functions $-\log [\boldsymbol{\alpha}]$ and $-\log [\mathbf{y}\mid\boldsymbol{\alpha}]$ \citep{holmes2017assigning}. The composite weight $w$ balances the relative contribution of the prior and likelihood of each observation in determining the weighted posterior $[\boldsymbol{\alpha}\mid\mathbf{y}]_w$ \citep{grunwald2017inconsistency}, where $w=1$ corresponds to standard Bayes and $w=0$ implies that the posterior is equivalent to the prior distribution. The weight typically is restricted to the range of $[0,1]$, which resembles the tempering of distributions used in complex MCMC schemes to improve mixing \citep{miller2018robust}. However, $w$ could be larger than 1, in which case the posterior increasingly relies on the data rather than the prior. Usually, $w$ is chosen \textit{a priori} or is selected using model selection techniques, though principled estimation frameworks have been proposed \citep{holmes2017assigning}. We use composite weights in a Bayesian context, although non-Bayesian methods have been developed that appropriately weight the score function at each observation in the maximum likelihood score equation \citep{majumder2021statistical}; the weight determines the compatibility of each observation with the model given the other observations, allowing for more robust estimation of parameters in the presence of outliers.

% homogeneous to heterogeneous powers
Conventional composite weights are specified to be homogeneous across observations. However, we specify the composite weights $w_{ij}$ such that the dyadic response $\widetilde{y}_{ij}$ is coarsened by a function of node-level covariates. 
We let $[\widetilde{y}_{ij}\mid\boldsymbol{\psi}]$ denote an unweighted dyadic data model with parameters $\boldsymbol{\psi}$ and $[\widetilde{y}_{ij}\mid\boldsymbol{\psi}]^{w_{ij}}$ denote the weighted data model constructed by powering $[\widetilde{y}_{ij}\mid\boldsymbol{\psi}]$ by $w_{ij}$. We normalize $[\widetilde{y}_{ij}\mid\boldsymbol{\psi}]^{w_{ij}}$ so that it integrates to one:
\begin{equation}\label{eq:npl}
    [\widetilde{y}_{ij}\mid\boldsymbol{\psi},w_{ij}] = \frac{[\widetilde{y}_{ij}\mid\boldsymbol{\psi}]^{w_{ij}}}{\int_\mathcal{Y}[\widetilde{y}_{ij}\mid\boldsymbol{\psi}]^{w_{ij}}d\widetilde{y}_{ij}},
\end{equation}
where $\mathcal{Y}$ is the support for $\widetilde{y}_{ij}$. We use the normalized weighted data model (\ref{eq:npl}) to attenuate the contribution of each edge for posterior estimation of $\boldsymbol{\psi}$.

We consider a normal data model in the case studies that follow because conventional dyadic models assume Gaussian noise.
We let $\widetilde{y}_{ij}$ be normally distributed with mean $\mathbf{\widetilde{x}}_{ij}'\boldsymbol{\beta}$ and variance $\sigma^2_y$, where $\mathbf{\widetilde{x}}_{ij}=\mathbf{x}_j-\mathbf{x}_i$ are differenced node-level covariates, $\boldsymbol{\beta}$ are regression coefficients, and $\boldsymbol{\psi}=\{\boldsymbol{\beta},\sigma^2\}$; this normal distribution can be expressed as an exponential family
\begin{equation}\label{eq:dm}
    [\widetilde{y}_{ij}\mid\boldsymbol{\psi}]=h(\widetilde{y}_{ij})g(\boldsymbol{\psi})\exp\{\boldsymbol{\eta}(\boldsymbol{\psi}) T(\widetilde{y}_{ij})\},
\end{equation}
with base measure $h(\widetilde{y}_{ij})\propto1$, normalizing constant $g(\boldsymbol{\psi})=(2\pi\sigma^2_y)^{-1/2}\exp\{-\boldsymbol{\beta}'\widetilde{\mathbf{x}}_{ij}\widetilde{\mathbf{x}}_{ij}'\boldsymbol{\beta}/(2\sigma^2)\}$, natural parameters $\boldsymbol{\eta}(\boldsymbol{\psi})=(\mathbf{\widetilde{x}}_{ij}'\boldsymbol{\beta}/\sigma^2_y,  -1/(2\sigma^2_y))'$, and sufficient statistics $T(\widetilde{y}_{ij})=\{\widetilde{y}_{ij}, \widetilde{y}_{ij}^2\}$. When (\ref{eq:dm}) is powered by $w_{ij}$, we have the weighted data model
\begin{equation}\label{eq:wdm}
    [\widetilde{y}_{ij}\mid\boldsymbol{\psi}]^{w_{ij}} \propto \exp\{w_{ij}\boldsymbol{\eta}(\boldsymbol{\psi})T(\widetilde{y}_{ij})\}
\end{equation}
with scaled natural parameters
\begin{equation}\label{eq:snp}
    w_{ij}\boldsymbol{\eta}(\boldsymbol{\psi})=\left(\frac{\mathbf{\widetilde{x}}_{ij}'\boldsymbol{\beta}}{(\sigma^2_y/w_{ij})},  -\frac{1}{2(\sigma^2_y/w_{ij})}\right)',
\end{equation}
where the composite weight $w_{ij}$ is grouped with the original variance parameter $\sigma^2_y$; this implies that the weighted data model (\ref{eq:wdm}) is the kernel of a Gaussian distribution with the same mean as the unweighted data model (i.e., $\mathbf{\widetilde{x}}_{ij}'\boldsymbol{\beta}$) and the updated heterogeneous variance $\sigma^2_y/w_{ij}$. Thus, the normalized weighted data model is Gaussian with mean $\mathbf{\widetilde{x}}_{ij}'\boldsymbol{\beta}$ and variance $\sigma^2_y/w_{ij}$.
In general, the normalized weighted data model $[\widetilde{y}_{ij}\mid\boldsymbol{\psi},w_{ij}]$ belongs to the same family as the unweighted data model if $[\widetilde{y}_{ij}\mid\boldsymbol{\psi}]$ is in the exponential family with base measure $h(\widetilde{y}_{ij})\propto 1$ and $w_{ij}$ can be grouped with the same set of parameters in each term of $\boldsymbol{\eta}(\boldsymbol{\psi})$ (i.e., as in (\ref{eq:snp})).

We aim to relate spatial and temporal proximity to the composite weights because we want dyads with large spatial or temporal lags to contribute less to posterior inference. Therefore, we propose the positive function
\begin{equation}\label{eq:hw}
    w_{ij}=\exp\{-\boldsymbol{\nu}(\mathbf{d}_{ij})'\boldsymbol{\gamma}\},
\end{equation}
where $\boldsymbol{\nu}(\mathbf{d}_{ij})$ are $q$ basis functions with basis coefficients $\boldsymbol{\gamma}$ and $\mathbf{d}_{ij}=(ds_{ij},dt_{ij})'$ represent the spatial lag $ds_{ij}$ and temporal lag $dt_{ij}$ between nodes $i$ and $j$. 
Although we focus on spatio-temporal dependencies, $\mathbf{d}_{ij}$ may contain node-level data in any parametric space. 
We consider monotonically increasing basis functions for $\boldsymbol{\nu}(\mathbf{d}_{ij})$ and constrain $\gamma_p>0, \; p=1,\ldots,q$, because we assume that observations are decreasingly dependent as $ds_{ij}$ and $dt_{ij}$ increase. 
When defining the weights as in (\ref{eq:hw}), the variance of $\widetilde{y}_{ij}$ under the normalized weighted data model is the power function $\sigma^2_y\exp\{\boldsymbol{\nu}(\mathbf{d}_{ij})'\boldsymbol{\gamma}\}$, which results in identifiable $\sigma^2_y$ and $\boldsymbol{\gamma}$. Thus, $\boldsymbol{\gamma}$ can be estimated using standard Bayesian computational methods (i.e., MCMC).

\subsection{Composite Weights \& Regularization}
The variance of the normalized weighted Gaussian data model increases with the spatial and temporal lags, and $\sigma^2_y$ is the unstructured error variance for the dyadic regression. Thus, the composite weighting scheme shrinks globally (via $\sigma^2_y$) and acts locally (via $\boldsymbol{\gamma}$). The composite weighting scheme resembles regularization techniques, such as the group lasso \citep{simon2013sparse} and the horseshoe prior \citep{carvalho2009handling}. Standard regularization techniques aim to reduce model complexity to better learn the inherent structure in the data. The proposed composite weighting scheme is similar to regularization but imposes heavier tails on the data model. 

Without composite weights, the mechanism may be obscured because the construction of superfluous edges in the fully-connected network may affect posterior inference on the mean structure. The composite weights allow the model to isolate the dyadic outcomes (the edges) that contribute most to learning the mechanistic dependence by discovering the optimal scale of dependence in space and time. A brief demonstration of the effect of composite weights on posterior inference is provided in Web Appendix A, where we fit a mechanistic population dynamic model to simulated count data \citep{schwob2023dynamic}. 

\subsection{Dyadic and Spatial Dependence}
Dyadic outcomes may exhibit dyadic and spatial dependence. For example, $\widetilde{y}_{ij}$ and $\widetilde{y}_{ik}$ share a common node and are likely correlated \citep{lindgren2010dyadic}. Additionally, nodes that are close in geographic space are likely dependent. To account for these sources of dependence, we consider the unweighted dyadic data model
\begin{equation}\label{eq:standarddyadic}
    [\widetilde{y}_{ij}\mid\boldsymbol{\beta},\widetilde{\eta}_{ij},\theta_i,\theta_j,\sigma^2_y] \equiv \text{N}\left(\widetilde{\mathbf{x}}_{ij}'\boldsymbol{\beta} + \widetilde{\eta}_{ij} + \theta_i + \theta_j,\sigma^2_y\right),
\end{equation}
where $\widetilde{\eta}_{ij}=\eta(\mathbf{s}_j)-\eta(\mathbf{s}_i)$ are differenced spatial random effects, $\boldsymbol{\eta}=(\eta(\mathbf{s}_1),...,\eta(\mathbf{s}_m))'$ are latent spatial random effects at the $m$ distinct node locations $\{\mathbf{s}_i\}$, and $\boldsymbol{\theta}\equiv(\theta_1,...,\theta_n)'$ are latent node-level random effects for the $n$ nodes. If each node has a distinct location, then $m=n$. The form of regression model in (\ref{eq:standarddyadic}) is conventional when accounting for dyadic and spatial dependence \citep{warren2023spatial} and extends multiple matrix regression \citep{wang2013examining}. The potential function for the unweighted data model (\ref{eq:standarddyadic}) is $\rho(\mathbf{x}(\mathbf{s}_i))=\mathbf{x}(\mathbf{s}_i)'\boldsymbol{\beta}+\eta(\mathbf{s}_i)$.

The random effects $\boldsymbol{\eta}$ and $\boldsymbol{\theta}$ account for latent attributes at the location-level and node-level, respectively, which induces correlation among edges that share a common location or node \citep{graham2020dyadic}. The spatial random effects $\boldsymbol{\eta}$ are differenced because they attempt to capture the latent difference between nodes that are not captured through $\widetilde{\mathbf{x}}_{ij}'\boldsymbol{\beta}$. In contrast, the latent node-level random effects $\boldsymbol{\theta}$ account for dyadic dependence between edges that share a common node. 
In dyadic modeling, the dependence between edges $\widetilde{y}_{ij}$ and $\widetilde{y}_{ik}$ attributed to the sharing of node $i$ is equivalent to the dependence between edges $\widetilde{y}_{ij}$ and $\widetilde{y}_{li}$ attributed to sharing node $i$ \citep{graham2020dyadic}; the ordered position that the common node takes does not affect the dyadic dependence among edges, so the $\boldsymbol{\theta}$ are added rather than subtracted in (\ref{eq:standarddyadic}).

% specification for eta
We specify a spatial Gaussian process for the prior $\boldsymbol{\eta}\sim \text{N}(\boldsymbol{0},\sigma^2_\eta \mathbf{R}(\phi))$, where $\mathbf{R}(\phi)_{ij}=\exp(-ds_{ij}/\phi)$ is an exponential-decay covariance function with spatial range $\phi$ and geodesic distance $ds_{ij}$ between nodes $i$ and $j$ \citep{cressie1991statistics}. In our case study, we specify an informative gamma prior for $\phi$ based on previous landscape genetics analyses. We propose values for $\phi$ in discrete support $\boldsymbol{\Phi}$, allowing us to compute a finite number of matrices $\mathbf{R}(\phi)$ prior to fitting the model. We used conventional conjugate priors for $\sigma^2_y$, $\boldsymbol{\beta}$, and $\boldsymbol{\theta}$ to facilitate efficient computation. The full Bayesian hierarchical model is provided in Web Appendix B and was implemented in Julia \citep{bezanson2017julia}. 

\subsection{Spatial Confounding}
% orthogonalization comment
Spatial confounding may be present in spatial generalized linear mixed models when both the spatial covariates and latent effects are smooth \citep{hodges2010adding,hughes2013dimension}. Thus, we constrain the spatial random effects to be orthogonal to the spatial fixed effects using conditioning by kriging in an MCMC algorithm \citep{rue2005gaussian}. 
We let $\mathbf{K}$ denote the $N\times m$ mapping matrix for $\boldsymbol{\eta}$ in the multivariate version of the unweighted data model (\ref{eq:standarddyadic}) and constrain $\boldsymbol{\eta}$ to be orthogonal to $\mathbf{C}=(\mathbf{K}'\mathbf{K})^{-1}\mathbf{K}'\mathbf{\widetilde{X}}$, where $\widetilde{\mathbf{X}}=(\widetilde{\mathbf{x}}_{12}, \ldots, \widetilde{\mathbf{x}}_{(n-1)n})'$ comprises the pairwise difference in node-level spatial covariates for each dyad. Then, we compute the constrained spatial random effects
\[
    \boldsymbol{\eta}^* = \boldsymbol{\eta} - \boldsymbol{\Sigma}_\eta\mathbf{C}\left(\mathbf{C}' \boldsymbol{\Sigma}_\eta \mathbf{C}\right)^{-1}\mathbf{C}'\boldsymbol{\eta},
\]
where $\boldsymbol{\Sigma}_\eta$ is the full-conditional covariance matrix for $\boldsymbol{\eta}$. Finally, we adjust the posterior variance for $\boldsymbol{\beta}$ to reflect the possible collinearity between the fixed and random effects by sampling
\[
    \boldsymbol{\beta}^* \sim \text{N}\left(\boldsymbol{\beta}, (\mathbf{C}'\mathbf{C})^{-1}\mathbf{C}'\boldsymbol{\Sigma}_\eta\mathbf{C}(\mathbf{C}'\mathbf{C})^{-1}\right),
\]
where $\boldsymbol{\beta}$ is the original coefficient vector \citep{hanks2015restricted}.

\subsection{Potential Surface Estimation}

The potential surface is of interest when analyzing the flow of a process. We estimate potential at location $\mathbf{s}^*$ with the posterior mean
\[
    \text{E}(\rho(\mathbf{s}^*)\mid\mathbf{y})\approx\frac1{K}\sum^K_{k=1}\left(\mathbf{x}(\mathbf{s}^*)'\boldsymbol{\beta}^{(k)}+\eta^{(k)}(\mathbf{s}^*)\right),
\]
where the $(k)$ superscript denotes the $k$th MCMC iteration and $\eta^{(k)}(\mathbf{s}^*)$ maximizes
\[
    [\eta(\mathbf{s}^*)\mid\boldsymbol{\eta}^{(k)}]=\int\int[\eta(\mathbf{s}^*),\sigma_\eta^{2(k)},\phi^{(k)}\mid\boldsymbol{\eta}^{(k)}]d\sigma_\eta^{2(k)}d\phi^{(k)}.
\]
We approximate the potential surface by computing $\text{E}(\rho(\mathbf{s}^*)\mid\mathbf{y})$ for each $\mathbf{s}^*\in\mathcal{S}$, where $\mathcal{S}$ is a fine grid spanning the study domain.
We quantify uncertainty at location $\mathbf{s}^*$ with the posterior variance
\[
    \text{Var}(\rho(\mathbf{s}^*)\mid\mathbf{y}) \approx \frac{1}{K}\sum^K_{k=1}\left(\mathbf{x}(\mathbf{s}^*)'\boldsymbol{\beta}^{(k)}+\eta^{(k)}(\mathbf{s}^*) - \text{E}(\rho(\mathbf{s}^*)\mid\mathbf{y})\right)^2.
\]

\section{Human Migration in Bronze Age Europe}

% introduce aDNA case study
The increasing accessibility of ancient human DNA (aDNA) facilitates our understanding of human history \citep{mallick2023allen}. One facet of human history relates to changes in the use of space by ancient populations \citep{narasimhan2019formation}. Previous studies have related ancient population genetic structure to the surrounding environment \citep{novembre2008genes}. More recently, \cite{frachetti2017nomadic} and \cite{schmid2023estimating} inferred the mechanism that governed ancient human movement in the highland geography of the Silk Roads and Holocene western Eurasia, respectively. In landscape genetics, such mechanisms are referred to as migratory surfaces \citep{petkova2016visualizing}, which are represented by potential surfaces for gene flow in geographic space.

% our goal
Migratory surfaces depict the effect that geographic features had on gene flow and are fundamental to understanding how ancient humans interacted with and navigated the environment \citep{petkova2016visualizing}. Our analysis focused on the migratory surface that guided human movement in Bronze Age Europe. We analyzed $n=398$ sequenced human genomes from the Allen Ancient DNA Resource \citep{mallick2023allen}; each genome sequence was dated between 6000-4500 years before the present (BP), recovered in Europe, and sampled in the 1240k library \citep{rohland2022three}. The dataset included the sequenced genome, estimated date of death, and retrieval location.

\begin{figure}
    \centering
    \includegraphics[width=\linewidth]{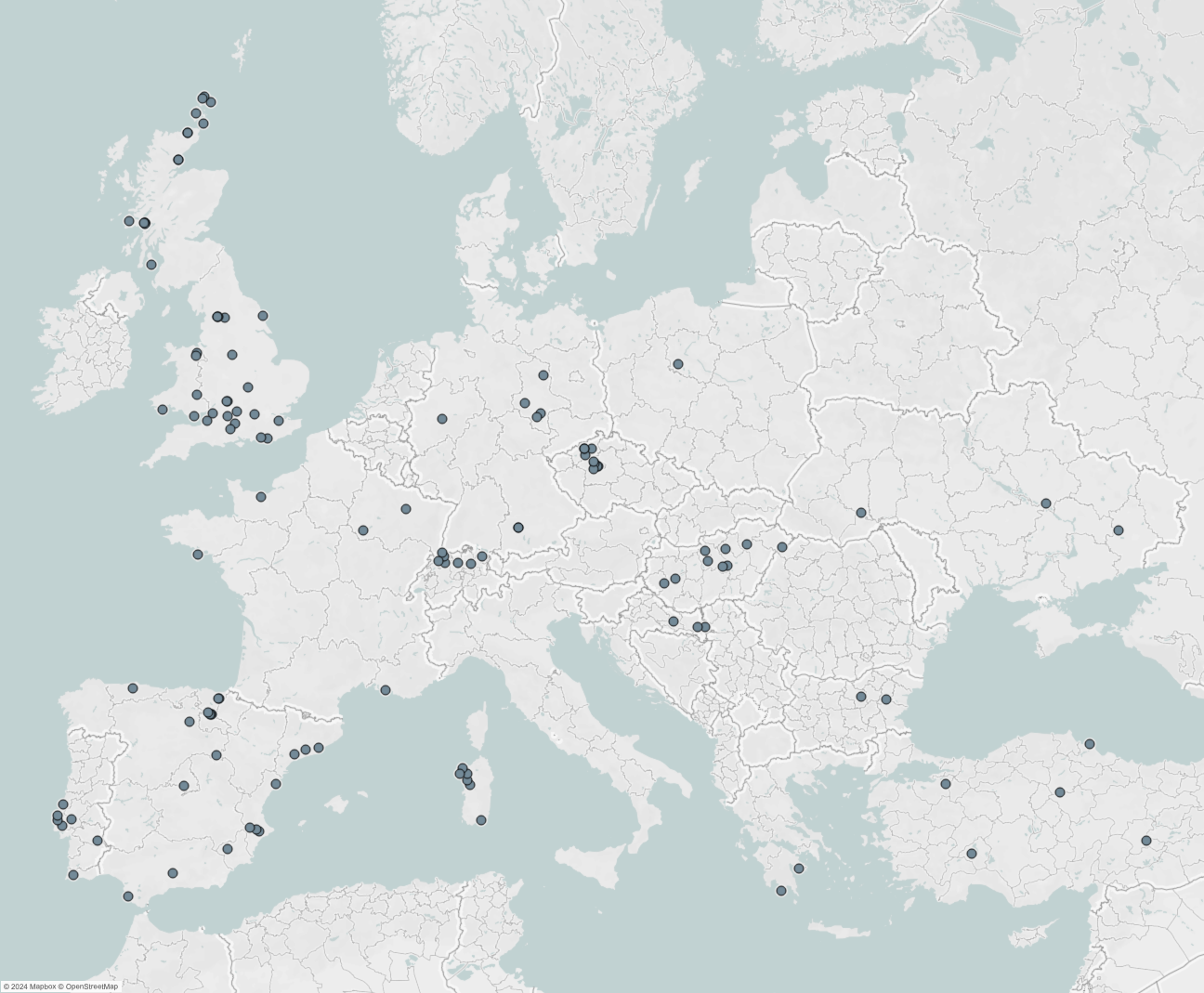}
    \caption{Locations of the analyzed sequenced human genomes from Bronze Age Europe.}
    \label{fig:aDNA_data}
\end{figure}

% discuss our data
We treated aDNA samples as spatially- and temporally-indexed nodes in a network.  Following standard practice in genomics, we projected the sequenced genomes on a 10-dimensional principal component space for dimension reduction to avoid shrinkage and obtain meaningful constraints for individuals with limited coverage \citep{patterson2006population}.
The dyadic outcome $\widetilde{y}_{ij}$ was defined as the weighted Euclidean distance between the projected genome sequences for nodes $i$ and $j$ in the principal component space, where weights were computed as the variance explained by each principal component; these weights are not related to the composite weights used in the data model. 
The dyadic outcome $\widetilde{y}_{ij}$ represented the genetic dissimilarity between individuals $i$ and $j$. By regressing genetic dissimilarity on a pairwise difference in covariates, our potential surface described gene flow (i.e., a migratory surface).
We used latitude, longitude, elevation, and proximity to historic fresh water as the node-level covariates $\mathbf{x}_i$; we also considered the squared effects of latitude and longitude, as well as their interaction. Elevation was approximated using a 30 arc second global relief grid \citep{wessel2019generic}, and proximity to historic freshwater was computed as the distance between a node location and the centroid of the nearest historic lake \citep{kelso2010introducing}. Locations of the samples are depicted in Figure \ref{fig:aDNA_data}.

% what did we do with the data once in this format?
We used the normalized weighted dyadic regression (\ref{eq:npl}) based on the model in (\ref{eq:standarddyadic}) to estimate the migratory surface. We regressed genetic dissimilarity $\widetilde{y}_{ij}$ on the change in observed environmental covariates $\widetilde{\mathbf{x}}_{ij}$, the change in latent spatial random effects $\widetilde{\eta}_{ij}$, and the node-level intercepts $\theta_i$ and $\theta_j$. We used an informative prior $[\phi] \equiv \text{Gamma}(400,250)$ for the spatial range in the Gaussian process on $\boldsymbol{\eta}$; these hyperparameters were selected based on exploratory data analyses of ancient human movement during different eras. We specified the set $\boldsymbol{\Phi}=\{0,...,\max(\textbf{ds})/3\}$ with increments of 10 geodesic units, where $\textbf{ds}$ denotes the vector of pairwise geodesic distances between the observed genetic sequences.

Composite weights may significantly improve posterior inference when defined as a function of pairwise lags that affect the dependence among nodes. We defined the composite weights as a function of the spatial lag in geodesic meters ($ds_{ij}$) and temporal lag in years BP ($dt_{ij}$) between nodes, where we scaled the spatial and temporal lags between edges to be in the unit interval. We specified $\boldsymbol{\nu}(\mathbf{d}_{ij})=(dt_{ij},ds_{ij})'$ with identity functions for ease in interpretation:
\begin{equation}\label{eq:cw}
    w_{ij}=\exp\{-(\gamma_1dt_{ij}+\gamma_2ds_{ij})\}.
\end{equation}
As the spatial or temporal lag increases, $w_{ij}$ decreases exponentially and the heterogeneous variance in the weighted data model increases. Therefore, edges that connect nodes with large spatial and temporal lags have inflated variances and do not obscure the mechanism. The identity functions appeared to sufficiently capture spatio-temporal dependence without introducing more parameters into the heterogeneous variance. However, with larger data sets, more complex basis functions may be used without sacrificing identifiability in the parameters governing the heterogeneous variance.

We fit four competing models that only differ in their specification for $w_{ij}$: (i) without composite weights (i.e., $w_{ij}=1$); (ii) with the composite weights in (\ref{eq:cw}) defined only as a function of the temporal lag $dt_{ij}$ (i.e., $\gamma_2=0$); (iii) with the composite weights defined only as a function of the spatial lag $ds_{ij}$ (i.e., $\gamma_1=0$); and (iv) with the composite weights defined as a function of both $dt_{ij}$ and $ds_{ij}$. We specified the prior $[\gamma_i] \equiv\text{Gamma}(2,2)$ for $i=1,2$, and we proposed values for a Metropolis-Hastings update using a truncated random-walk on the positive domain. 
Other prior distributions for $\boldsymbol{\gamma}$ with support $\mathbb{R}^+$ may work; however, we found the gamma prior to provide the most efficiently mixed Markov chains.

We compared these models using continuous-rank probability scores (CRPS) to demonstrate how accounting for spatial and temporal proximity between nodes affects the model fit. CRPS is a scoring function that compares a continuous-valued variable ($\widetilde{y}_{ij}$) with its predicted distribution, taking into account the shape and location of the predicted distribution as opposed to a predicted point estimate \citep{matheson1976scoring, gneiting2007probabilistic}.

The first model resulted in the greatest (i.e., worst) CRPS $(3.037\times 10^{-3})$, likely due to its inability to account for dependence structures in space or time. Without heteroskedasticity, dyads contributed equally to the estimation of the parameters in the mean structure of the dyadic data model. Thus, edges connecting uncorrelated nodes contributed to the estimation of the potential surface as much as edges connecting correlated nodes.

The second and third models resulted in a CRPS of $2.994\times 10^{-3}$ and $3.016\times 10^{-3}$, respectively. Therefore, both models fit the observed data better than the model without composite weights. The lower CRPS for the second model implies that accounting for the dependence structure in time improved the model fit more than accounting for the dependence structure in space. Finally, the fourth model resulted in a CRPS of $3.013\times 10^{-3}$. The fourth model fit the data better than the first and third model due to the inclusion of temporal lags as an additional auxiliary data source. However, the model with only temporal lags outperformed the model with both temporal and spatial lags because the increased complexity in the fourth model resulted in higher heterogeneous variance, which was penalized in the CRPS.

The proposed method indicates that accounting for temporal dependence in the fully-connected network improves model fit. Spatial dependence likely contributed less to learning the migratory surface because ancient humans could migrate over the entire study domain within one or two generations \citep{patterson2012ancient}, which is rapid relative to the coarse temporal resolution of the data. Additionally, unlike the temporal domain, movement in the spatial domain is bidirectional, which may contribute to the complexity of spatial dependence in the data.

\begin{figure}
    \centering
    \begin{subfigure}[t]{0.49\textwidth}
        \centering
        \includegraphics[width=\linewidth]{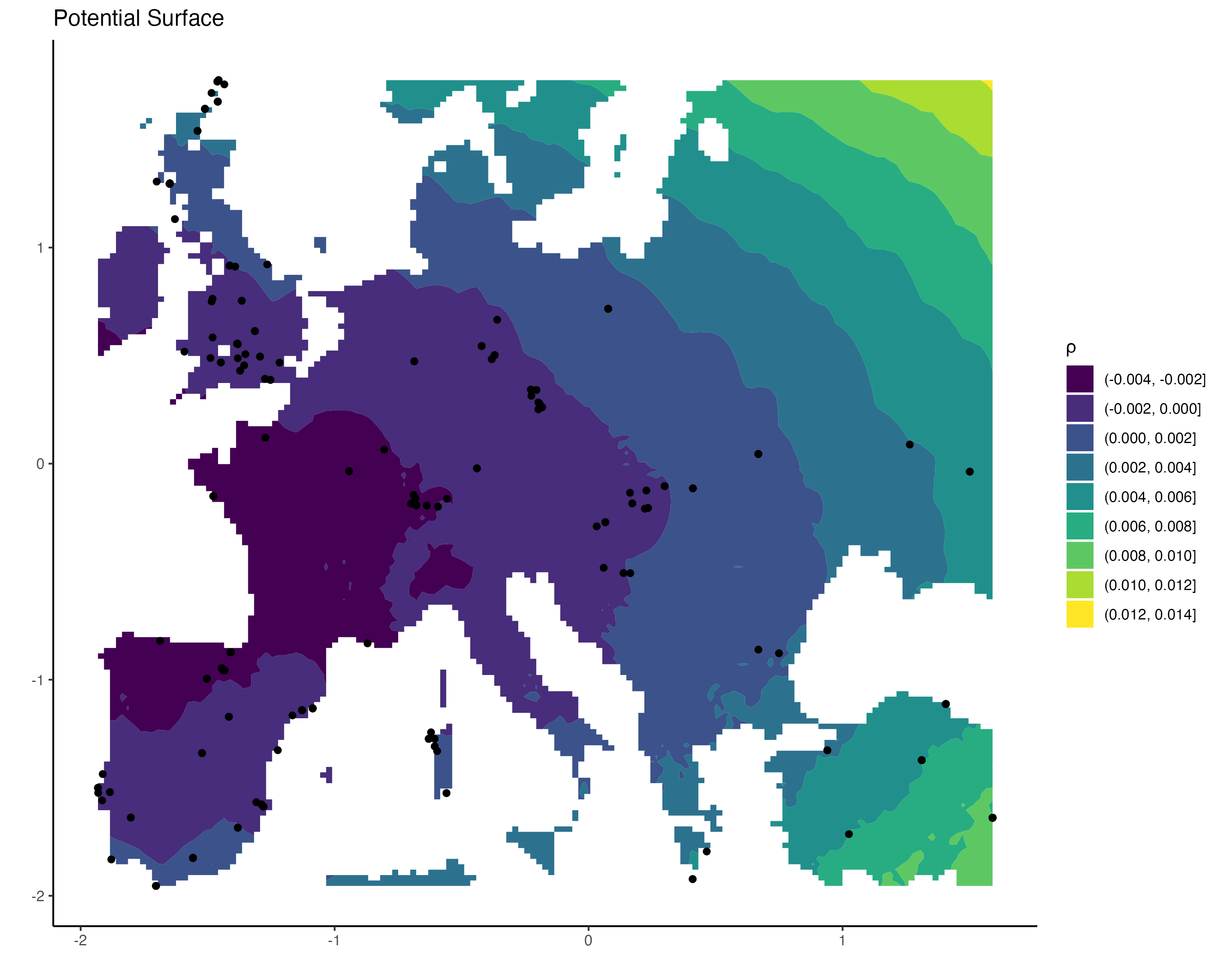}
    \end{subfigure}
    \hfill
    \begin{subfigure}[t]{0.49\textwidth}
        \centering
        \includegraphics[width=\linewidth]{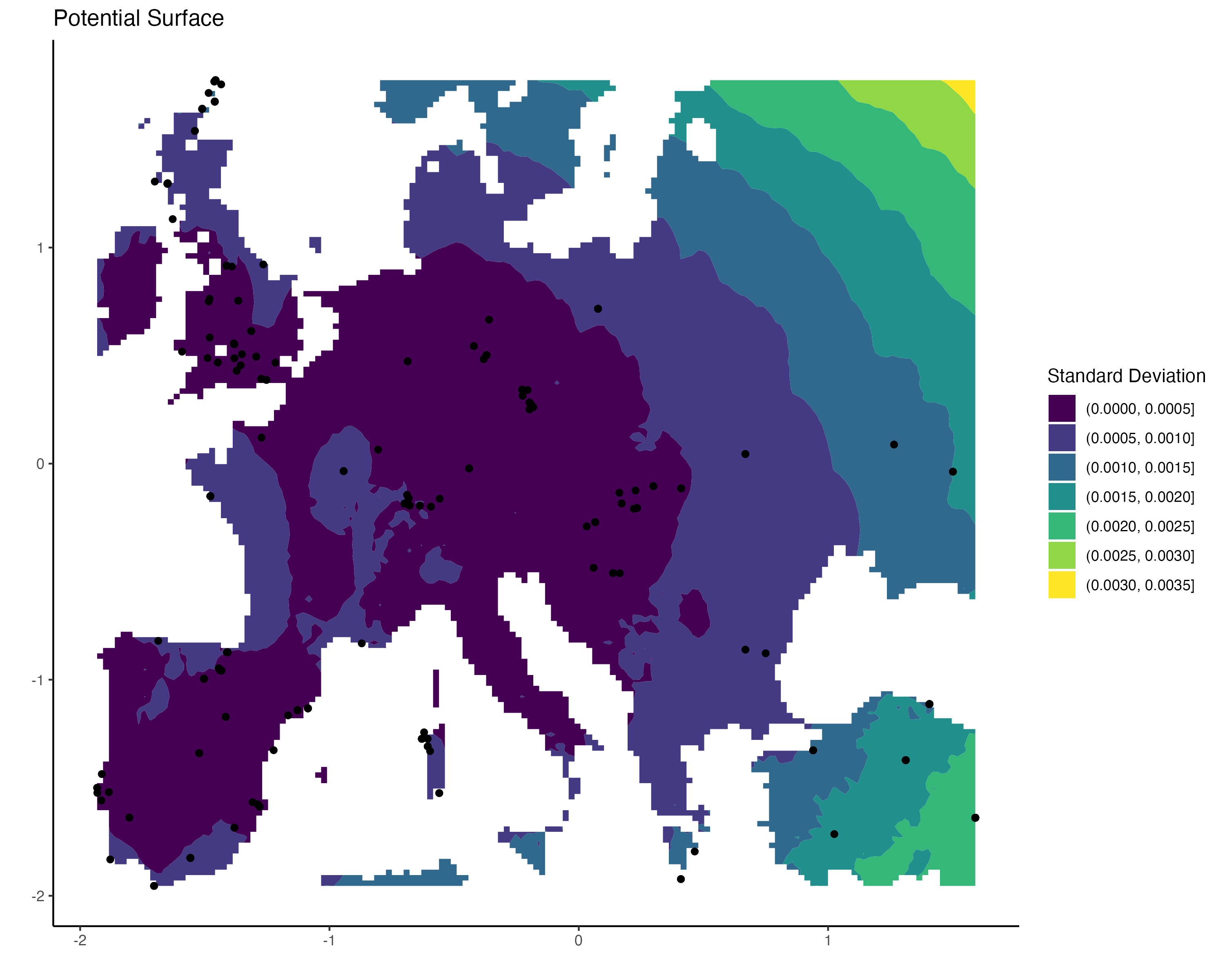}
    \end{subfigure}
    \caption{The inferred potential surface using posterior estimates from our Bayesian hierarchical dyadic model. (a) Posterior mean of the potential surface for ancient human gene flow in Bronze Age Europe from 6000-4500 BP. Barriers to gene flow are represented by high potential (yellow) regions. Regions with less potential were more conducive to gene flow. (b) Posterior standard deviation of the potential surface.}
    \label{fig:potPlot}
\end{figure}

% explain findings
We estimated the migratory surface using posterior mean estimates for $\boldsymbol{\beta}$ and $\boldsymbol{\eta}$ from the second model (see Figure \ref{fig:potPlot}). 
The higher the potential in a region, the more resistant that region is to gene flow (i.e., a barrier). Similarly, the lower potential regions were more conducive to gene flow (i.e., genes flowed more freely in these regions).
The inferred potential surface indicates that Central and Western Europe were most conducive to gene flow during the Bronze Age, whereas Eurasia was more resistant to gene flow; this supports the hypothesis that pastoralists migrated from the Yamnaya Steppe in Eurasia to Central and Western Europe during this time period \citep{narasimhan2019formation}.
The migratory surface also reveals that northern regions tended to be more resistant to gene flow than southern regions, which aligns with well-supported theories on ancient human movement throughout Europe \citep{patterson2012ancient}.
Additionally, the relatively high potential of the Swiss Alps compared to the surrounding region implies that this mountain range acted as a barrier to gene flow.
%Finally, coastal regions have lower potential relative to their surrounding areas, which supports the theory that ancient humans often navigated coasts during migration \citep{bellwood2014first}.

% discuss aDNA case study
Analysis of the aDNA data set was computationally feasible because our model scales well with larger data sets due to its formulation as dyadic regression; our MCMC algorithm required 171 minutes to perform 500,000 iterations (49it/s) on a 3.2 GHz processor with 64 GB of RAM. However, the coarseness of the data in space and time presented a challenge. Among the 398 nodes analyzed, the largest pairwise spatial and temporal lags were 508,558 geodesic meters and 1,472 years, respectively. Many edges connected nodes that were distant in space and time. With composite weights, the heterogeneous variances for such edges were larger, and these edges contributed less to the estimation of the potential surface. 

\section{Discussion}

% what we did
We developed a scalable Bayesian hierarchical framework that allowed us to infer spatio-temporal processes using dyadic regression and composite likelihoods. Dyadic regression is used to study the flow of processes across a network, yet most models do not account for the spatio-temporal structure of data despite the dynamics of the process occurring in space and time \citep{graham2020dyadic,warren2023spatial}. We linked these models with advection-diffusion differential equations to infer the mechanism that governs the structure of the data in space and time. 

% we used composite weights to help find subgraph structure
To use dyadic regression, we constructed a network from the spatio-temporal data. Without knowledge of the spatio-temporal process, it is difficult to determine which nodes to connect. Thus, we constructed a fully-connected network and mechanistically modeled the dependence in space and time between nodes using composite likelihoods. The composite weights attenuated the contribution of edges in the fully-connected network using auxiliary information, such as spatial and temporal lags between nodes. Advection-diffusion differential equations relate a temporal change in the process of interest to a change in the space that contains the process. We used composite weights to impose similar dependence in space and time to infer subgraph structures that aid in the estimation of the mechanism. 
%The composite weights allowed the model to identify the correct scale of dependence in space and time.

% discuss the imposition of dependence structures
In spatial statistics, there are several methods to impose dependence structures at the node-level \citep{cressie1991statistics}. Common spatio-temporal statistical models such as the (intrinsic) conditional autoregressive model put significant structure on the covariance when assuming that dependence between observations decreases as they grow distant in space and time. In contrast, dyadic regression places such dependence structures on the edges because they do not model node-level data directly \citep{graham2020dyadic}. We extended dyadic regression to induce dependence structures common in spatio-temporal statistical models.

% composite likelihoods don't cost us computational efficiency
Computational efficiency was a primary motivation in developing the proposed model. The use of composite weights maintained conjugacy in the model parameters because the normalized weighted data model belonged to the same family as the unweighted data model. Thus, any conjugacy imposed by the unweighted Gaussian data model remained after normalizing the weighted data model.

% Gaussian data model
For Gaussian data models, composite weights induce a heterogeneous variance that is a function of spatial and temporal lags. Heteroskedasticity is well-studied in space and time \citep{wikle2010general,cressie2015statistics}. 
We found that, when the composite weights are specified as in (\ref{eq:hw}), one or two basis functions were sufficient to adequately recover spatio-temporal dependence; as the number of parameters in the heterogeneous variance increases, it is difficult to adequately estimate the parameters.
As spatial and temporal lags increase, the normalized weighted data model variance increases, thereby constraining the model to focus on shorter resolutions in space and time where the process of interest actually occurs. Thus, the composite likelihoods weigh the contribution that each dyad has on posterior inference. A similar concept occurs in weighted least-squares regression, which assigns relatively less weight to ordinates of the empirical semivariogram corresponding to large lags in any parametric space \citep{cressie1985fitting}. 

% comparing our method with circuit theoretic method
We provide a conventional landscape genetics example in Web Appendix C, where we inferred the mechanism that governed gene flow in \textit{Rupicapra rupicapra} populations in the Swiss alps using our dyadic regression model. The resulting inference aligns with that obtained by the circuit theoretic method outlined in \cite{hanks2013circuit}; this comparison demonstrates that our method can be used in a variety of settings and provides results similar to conventional landscape genetic methods. However, the \textit{R. rupicapra} case study is limited because the data set does not contain temporal indices. Thus, inference is limited to mechanisms governing the spatial gene flow in \textit{R. rupicapra} populations. 

% mechanism inference in other spaces
We focused on processes that occur in space and time due to the ubiquity of space-time dynamics in many fields of science. However, our proposed method can infer mechanisms in any parametric space when all nodes have the relevant auxiliary information. For example, potential surfaces outside of geographic space in our aDNA case study could be accommodated, such as mechanisms that affect cultural or demographic processes \citep{hubisz2020mapping}. It may be challenging to learn potential surfaces in such spaces due to a lack of node-level auxiliary information. However, in the field of ancient human movement, other data sources, such as archaeological data, can help infer potential surfaces in other parametric spaces, such as technological or cultural practices \citep{becerra2020timing}.

% Paragraph for "different levels of differencing"
Our use of composite weights is applicable to any level of change. We demonstrated the use of composite likelihoods for first-order change, which is the difference between nodes. Applications that are concerned with acceleration through nodes will require analyzing second-order differences, which is the difference between edges \citep{kolaczyk2014statistical}; such a data set would have size $n(n-1)(n-2)/2 > N$. As the order of difference increases, the artificially constructed data set increases in size. Thus, a principled model-based quantity to attenuate the contribution of each differenced outcome may be helpful in inferring the latent mechanism of interest. Conversely, some studies may be interested in aggregating the observed data set, where collections of nodes are treated as a single unit, or a hyper-node \citep{schwob2019modeling}; in such a case, the transformed data set decreases in size. Although first-order changes are most commonly studied, higher-order changes may also be considered.

\section*{Acknowledgements}

This research was supported by the NSF Graduate Research Fellowship Program.

\section*{Data Availability}

The aDNA dataset is publicly available from the Allen Ancient DNA Resource \citep{mallick2023allen}. Code used for data analysis in this manuscript and the Appendices is publicly available at: \\ https://github.com/michaelschwob/composite-dyadic-models. 

%%%%%%%%%%%%%%%%%%%%%%%%%%%%%%%%%%%%%%%%%%%%%%%%%%%%%%% qwerty 

%\appendix

\section*{Appendix A}
We demonstrate the effect of heterogeneous composite weights on posterior inference with a simulated autoregressive process
\begin{equation}\label{eq:ar}
    y_t \sim \text{N}(y_{t-1} +(\mathbf{x}_t - \mathbf{x}_{t-1})'\boldsymbol{\beta},\sigma^2_0),
\end{equation}
where $\mathbf{x}_t=(x_{t,1}, x_{t,2})'$ are covariates for observation $t$, $\boldsymbol{\beta}=(1.3, 0.8)'$ are regression coefficients, $\sigma^2_0=0.01$ is the variance of the difference in successive observations, and $t\in\mathcal{T}\equiv\{1,...,15\}$ following conventional discrete time series notation. We initialized the time series with $y_0=0$ and $\mathbf{x}_0=(0,0)'$ and simulated $x_{t,j}=x_{t-1,j} + \epsilon_{t,j}$ for $j=1,2$, where $\epsilon_{t,j}\sim\text{N}(0, 0.5^2)$. We note that (\ref{eq:ar}) may be interpreted as a mechanistic population dynamic model, where $y_t$ denotes the abundance of a species on day $t$ and $\mathbf{x}_t$ are the environmental covariates on day $t$ \citep{schwob2023dynamic}. The data model (\ref{eq:ar}) is equivalent to $y_{(t-1)t} \sim \text{N}(\widetilde{\mathbf{x}}_{(t-1)t}'\boldsymbol{\beta},\sigma^2_0),$ where $y_{(t-1)t}=y_t-y_{t-1}$ and $\widetilde{\mathbf{x}}_{(t-1)t}=\mathbf{x}_t-\mathbf{x}_{t-1}$.
We assumed a lack of knowledge about the AR(1) data generation and aimed to recover the true potential surface (which governs population growth in the population dynamic context) and the autoregressive dependence, where $\mathcal{T}$ is treated as a one-dimensional discrete space. 

We constructed a fully-connected network comprising nodes at locations $\mathcal{T}$ and edge weights defined as the difference in the node-level responses:
\begin{equation}\label{eq:basic}
    y_{ij} \sim \text{N}(\widetilde{\mathbf{x}}_{ij}'\boldsymbol{\beta},\sigma^2),
\end{equation}
where $i,j\in\mathcal{T}$ and $j>i$. We note that (\ref{eq:basic}) is correctly specified when regressing only on edges where the change in time is $dt_{ij}=1$. Additionally, the variance parameter in (\ref{eq:basic}) is denoted $\sigma^2$ rather than $\sigma^2_0$ because it is the variance of all pairwise differences rather than the variance of successive observations.
The normalized weighted data model is
\begin{equation}\label{eq:nwd}
    [y_{ij}\mid\boldsymbol{\beta},\sigma^2,w_{ij}] \equiv \frac{[y_{ij}\mid\boldsymbol{\beta},\sigma^2]^{w_{ij}}}{\int_\mathcal{Y}[y_{ij}\mid\boldsymbol{\beta},\sigma^2]^{w_{ij}}dy_{ij}},
\end{equation}
where $[y_{ij}\mid\boldsymbol{\beta},\sigma^2]$ is the unweighted data model in (\ref{eq:basic}).

We considered two scenarios to demonstrate the effect that composite weights had on inferring the potential surface. First, we regressed the fully-connected network without composite weights; this was equivalent to defining $w_{ij}=1$ for each edge in (\ref{eq:nwd}). Second, we regressed the fully-connected network with composite weights $w_{ij}=\exp\{-\boldsymbol{\nu}(dt_{ij})'\boldsymbol{\gamma}\}$, where $\nu_p(dt_{ij})=dt_{ij}^{p/3}$ for $p=1,...,6$; this specification implies that as $dt_{ij}$ increased, the dependence between nodes $i$ and $j$ decreased. In both scenarios, we considered the conjugate priors $\boldsymbol{\beta}\sim\text{N}(\boldsymbol{0}, 10^6\mathbf{I})$ and $\sigma^2\sim\text{IG}(0.01,0.01)$. In the second scenario, we considered a $\text{Gamma}(10, 2)$ prior for each $\gamma_p$ and proposed values for a Metropolis-Hastings update with an adaptively-tuned random-walk \citep{roberts2009examples}. 

We obtained the potential surface for both scenarios by evaluating the potential function $\rho_t=\mathbf{x}_t'\boldsymbol{\beta}$ for $t \in \mathcal{T}$ using the posterior mean of $\boldsymbol{\beta}$; additionally, we used the 2.5th and 97.5th percentiles of $\boldsymbol{\beta}$ to quantify uncertainty. The output is provided in Web Figure 1.
The inferred potential surface from the first scenario fails to capture the true potential for every point in $\mathcal{T}$, whereas the inferred potential surface from the second scenario captures the true potential for every point in $\mathcal{T}$.
The first scenario contained a homogeneous variance and, thereby, assumed that all edges contributed equally to the learning of $\rho_t=\mathbf{x}_t'\boldsymbol{\beta}$. However, based on the true AR(1) data generation, we know that observations farther in the space $\mathcal{T}$ are less dependent, and edges connecting such observations will likely obscure the mechanism. 
In the second scenario, the composite weights reduced the contribution that edges connecting nodes distant in $\mathcal{T}$ had on posterior inference for $\boldsymbol{\beta}$ and the potential function. Thus, the second scenario better predicted the potential surface across the entire domain $\mathcal{T}$.

\begin{figure}
    \centering
    \includegraphics[width=\linewidth]{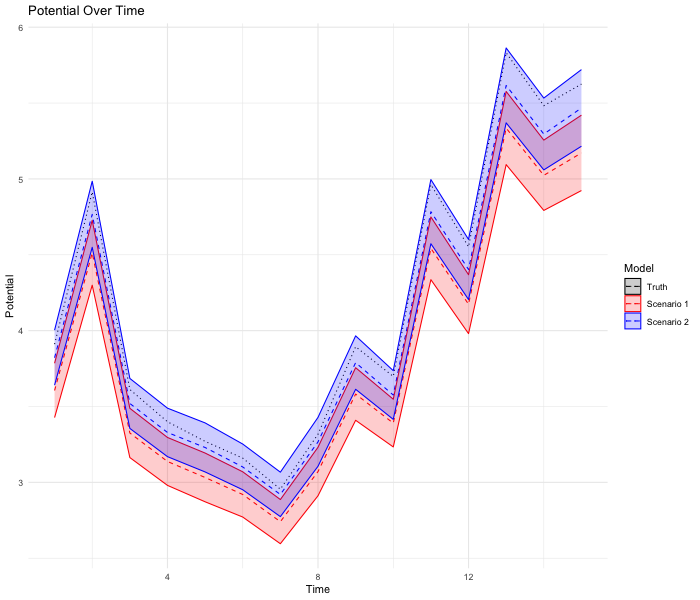}
    \caption*{\textbf{Web Figure 1:} Posterior inference for the potential surface across $\mathcal{T}$. The true potential is denoted by a dotted black line. Posterior means and the 95\% credible intervals for the potential surface are provided for the first (red) and second (blue) scenarios.}
    \label{fig:potPlot_ar1}
\end{figure}

The inferred composite weights imposed a dependence structure in time that resembled the dependence found in an autoregressive process. Posterior inference for the composite weights $w_{ij}=\exp(-\boldsymbol{\nu}(dt_{ij})'\boldsymbol{\gamma})$ is depicted in Web Figure 2, which indicates that the heterogeneous variance increased as $dt_{ij}$ increased. For example, when $dt_{ij}=1$, the variance for $y_{ij}$ was approximately $2\sigma^2$ and, when $dt_{ij}=3$, the variance was approximately $5\sigma^2$.

\begin{figure}
    \centering
    \includegraphics[width=\linewidth]{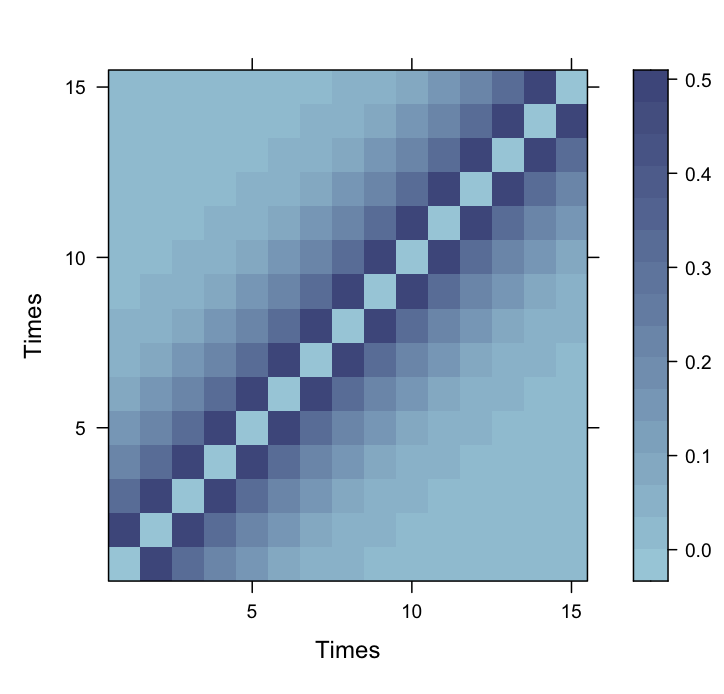}
    \caption*{\textbf{Web Figure 2:} Posterior composite weights per difference in time. The color for each cell represents the inferred composite weight for an edge connecting a node at time $t_r$ and $t_c$, where $t_r$ and $t_c$ denote the time assigned to the rows and columns, respectively. The values in the diagonal are zero because we do not regress edges with $dt=0$.}
    \label{fig:imagePlot}
\end{figure}

% discuss AR(1) case study
We included the first scenario to demonstrate that regressing over a fully-connected network without composite weights may result in poor inference. Without knowledge of the temporal process, it may be difficult to determine which edges to construct in a network; this difficulty is exacerbated when inferring spatio-temporal processes. Thus, we constructed a fully-connected network and used composite weights to leverage temporal information that helped us account for mechanistic dependence. The composite weights enabled the model to isolate the dyadic outcomes that contributed the most to learning the mechanistic dependence (i.e., those close in time).

\newpage
\section*{Appendix B}

We fit the following Bayesian hierarchical model in our aDNA case study:

\begin{align*}
    [y_{ij}\mid\boldsymbol{\beta},\widetilde{\eta}_{ij},\theta_i,\theta_j,\sigma^2_y,w_{ij}] &= \frac{[y_{ij}\mid\boldsymbol{\beta},\widetilde{\eta}_{ij},\theta_i,\theta_j,\sigma^2_y]^{w_{ij}}}{\int_\mathcal{Y}[y_{ij}\mid\boldsymbol{\beta},\widetilde{\eta}_{ij},\theta_i,\theta_j,\sigma^2_y]^{w_{ij}}dy_{ij}},\\
    \boldsymbol{\beta} &\sim \text{N}(\boldsymbol{0},10^6\mathbf{I}),\\
    \boldsymbol{\eta} &\sim \text{N}(\boldsymbol{0}, \sigma^2_\eta \mathbf{R}(\phi)),\\
    \boldsymbol{\theta} &\sim \text{N}(\boldsymbol{0},\sigma^2_\theta \mathbf{I}),\\
    \sigma^2_y &\sim \text{IG}(0.01, 0.01),\\
    \sigma^2_\eta &\sim \text{IG}(0.01, 0.01),\\
    \sigma^2_\theta &\sim \text{IG}(0.01, 0.01),\\
    \phi &\sim \text{Gamma}(400,250),\\
    w_{ij} &= \exp\left(-(\gamma_1dt_{ij} + \gamma_2ds_{ij})\right),\\
    \gamma_k &\sim \text{Gamma}(2, 2),
\end{align*}
where $i,j\in\{1,...,n\}$, $k\in\{1,2\}$, and 
%$\boldsymbol{\Phi}=\{0,...,\max(\mathbf{ds})/4\}$ with increments of 10 geodesic units, and
\begin{equation*}
    [y_{ij}\mid\boldsymbol{\beta},\widetilde{\eta}_{ij},\theta_i,\theta_j,\sigma^2_y]\equiv\text{N}\left(\mathbf{\widetilde{x}}_{ij}'\boldsymbol{\beta} + \widetilde{\eta}_{ij} + \theta_i + \theta_j ,\sigma^2_y\right).
\end{equation*}

\clearpage

\section*{Appendix C}

We apply the proposed methods to study spatial gene flow in alpine chamois (\textit{Rupicapra rupicapra}) to compare the results of our method with those found in \cite{hanks2013circuit}. The alpine chamois is a mountain ungulate native to Europe and is a conserved species in France. The \textit{adegenet} R package described in \cite{jombart2008adegenet} contains microsatellite allele data from nine loci for 335 individual chamois. Each genetic sequence has a spatial reference with elevation data on a $104\times80$ grid, where each grid cell has an area of 40km$^2$. More information on the data can be found in \cite{jombart2008adegenet}.

\cite{hanks2013circuit} examined the effect of elevation on gene flow in the alpine chamois population in the Bauges mountains of France. They found a strong nonlinear relationship between elevation and gene flow, with alpine chamois preferring habitats that are moderately high in elevation. Thus, we used elevation and squared elevation as the node-level covariates. For the dyadic outcomes, we computed the Manhattan distance between individuals' genomes using the \textit{adegenet} package. 

We fit the Bayesian hierarchical model in Web Appendix B with two modifications. First, we used the informative prior $[\phi] \equiv \text{Gamma}(500,5)$ for the spatial range parameter based on inference from \cite{jombart2008adegenet}. Second, we fixed $\gamma_1=0$ because the data set does not contain temporal indices per observation. The resulting posterior $\mathbf{x}'\boldsymbol{\beta}$ surface in Web Figure 3a aligns with the inference provided in Figure 3c of \cite{hanks2013circuit}. However, the posterior latent $\boldsymbol{\eta}$ surface obscures the $\mathbf{x}'\boldsymbol{\beta}$ surface when computing the potential surface (see Web Figure 3b); this suggests that there are fine-scale processes governing gene flow in the region that were not recovered by the circuit theoretic method. Web Figure 4 depicts that, on average, regions in the Bauges mountains with elevation slightly greater than 1500m offer the lowest potential (highest conductance) to gene flow in the chamois population; Figure 3b in \cite{hanks2013circuit} supports this finding.

\begin{figure}
    \centering
    \begin{subfigure}{0.48\textwidth}
        \includegraphics[width=\linewidth]{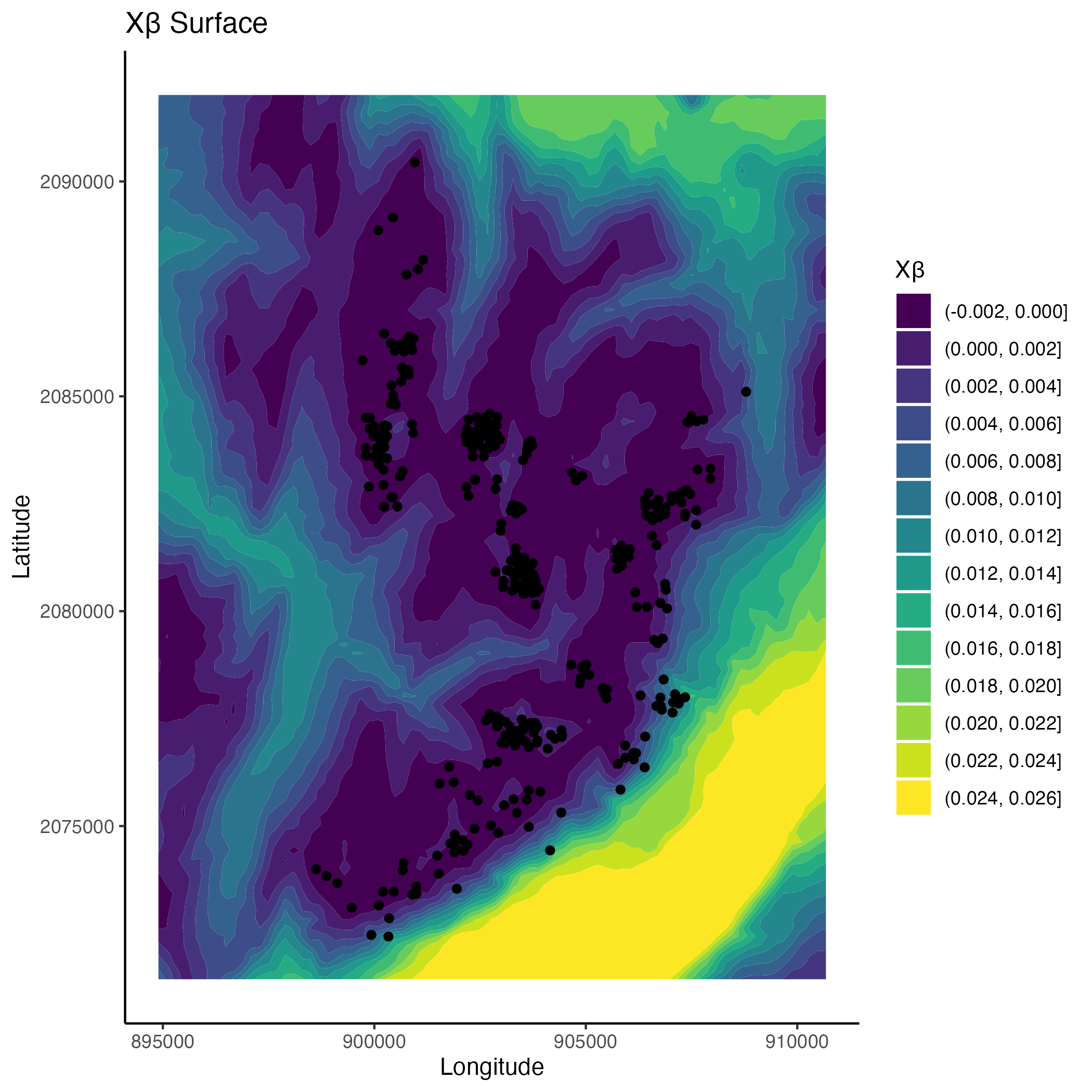}
        \caption{}\label{fig:1a}
      \end{subfigure}
    \begin{subfigure}{0.48\textwidth}
        \includegraphics[width=\linewidth]{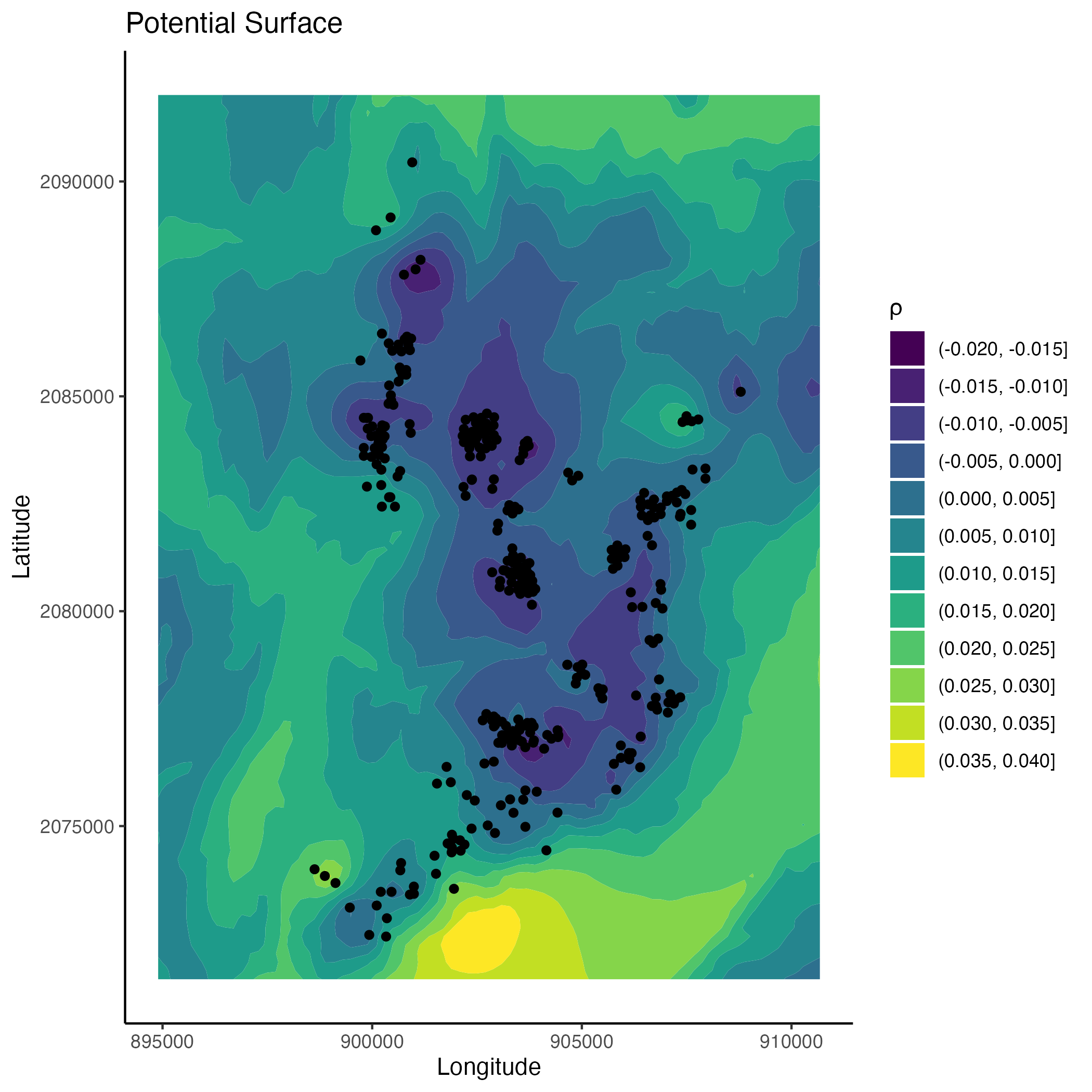}
        \caption{}\label{fig:1b}
      \end{subfigure}
      \caption*{\textbf{Web Figure 3:} (a) The posterior $\mathbf{x}'\boldsymbol{\beta}$ surface for \textit{R. rupicapra} gene flow in the Bauges mountains of France. Regions with lower values contribute less to the potential surface, indicating they offer more conductance to gene flow. (b) The posterior potential surface. Regions with high potential correspond to regions with high resistance in the circuit theoretic perspective. Similarly, regions with low potential correspond to regions with high conductance for gene flow.}
\end{figure}

\begin{figure}
    \centering
    \includegraphics[width=.5\linewidth]{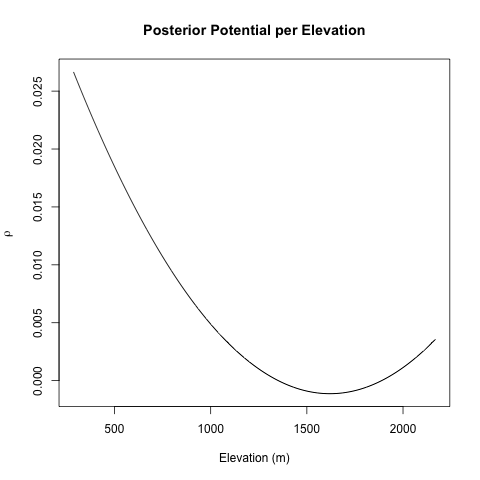}
    \caption*{\textbf{Web Figure 4:} The posterior potential at elevation. Elevations slightly over 1500m have the lowest potential, indicating they provide the highest conductance to gene flow in the chamois population.}
    \label{fig:rupi_elev}
\end{figure}

\clearpage

\bibliographystyle{abbrvnat}
\bibliography{bibliography.bib}

\begin{thebibliography}{59}
\providecommand{\natexlab}[1]{#1}
\providecommand{\url}[1]{\texttt{#1}}
\expandafter\ifx\csname urlstyle\endcsname\relax
  \providecommand{\doi}[1]{doi: #1}\else
  \providecommand{\doi}{doi: \begingroup \urlstyle{rm}\Url}\fi

\bibitem[Becerra-Valdivia and Higham(2020)]{becerra2020timing}
L.~Becerra-Valdivia and T.~Higham.
\newblock The timing and effect of the earliest human arrivals in {N}orth {A}merica.
\newblock \emph{Nature}, 584\penalty0 (7819):\penalty0 93--97, 2020.

\bibitem[Berliner(2003)]{berliner2003physical}
L.~M. Berliner.
\newblock Physical-statistical modeling in geophysics.
\newblock \emph{Journal of Geophysical Research: Atmospheres}, 108\penalty0 (D24), 2003.

\bibitem[Berliner et~al.(2023)]{berliner2023excursions}
L.~M. Berliner et~al.
\newblock Excursions in the {B}ayesian treatment of model error.
\newblock \emph{PloS ONE}, 18\penalty0 (6):\penalty0 e0286624, 2023.

\bibitem[Bezanson et~al.(2017)]{bezanson2017julia}
J.~Bezanson et~al.
\newblock Julia: A fresh approach to numerical computing.
\newblock \emph{SIAM review}, 59\penalty0 (1):\penalty0 65--98, 2017.

\bibitem[Brillinger et~al.(2012)]{brillinger2012use}
D.~R. Brillinger et~al.
\newblock The use of potential functions in modelling animal movement.
\newblock \emph{Selected Works of David Brillinger}, pages 385--409, 2012.

\bibitem[Carvalho et~al.(2009)Carvalho, Polson, and Scott]{carvalho2009handling}
C.~M. Carvalho, N.~G. Polson, and J.~G. Scott.
\newblock Handling sparsity via the horseshoe.
\newblock In \emph{Artificial Intelligence and Statistics}, pages 73--80. PMLR, 2009.

\bibitem[Cressie(1985)]{cressie1985fitting}
N.~Cressie.
\newblock Fitting variogram models by weighted least squares.
\newblock \emph{Journal of the International Association for Mathematical Geology}, 17:\penalty0 563--586, 1985.

\bibitem[Cressie(1991)]{cressie1991statistics}
N.~Cressie.
\newblock \emph{Statistics for Spatial Data}.
\newblock John Wiley \& Sons, 1991.

\bibitem[Cressie and Wikle(2015)]{cressie2015statistics}
N.~Cressie and C.~K. Wikle.
\newblock \emph{Statistics for Spatio-Temporal Data}.
\newblock John Wiley \& Sons, 2015.

\bibitem[Daniela et~al.(2012)]{daniela2012air}
B.~Daniela et~al.
\newblock Air pollution steady-state advection-diffusion equation: The general three-dimensional solution.
\newblock \emph{Journal of Environmental Protection}, 2012, 2012.

\bibitem[Dunphy-Guzman et~al.(2006)Dunphy-Guzman, Finnegan, and Banfield]{dunphy2006influence}
K.~A. Dunphy-Guzman, M.~P. Finnegan, and J.~F. Banfield.
\newblock Influence of surface potential on aggregation and transport of titania nanoparticles.
\newblock \emph{Environmental Science \& Technology}, 40\penalty0 (24):\penalty0 7688--7693, 2006.

\bibitem[Fletcher et~al.(2023)Fletcher, Dillingham, and Parry]{fletcher2023simple}
D.~Fletcher, P.~W. Dillingham, and M.~Parry.
\newblock A simple and robust approach to {B}ayesian modelling of overdispersed data.
\newblock \emph{Environmental and Ecological Statistics}, pages 1--20, 2023.

\bibitem[Fosdick and Hoff(2015)]{fosdick2015testing}
B.~K. Fosdick and P.~D. Hoff.
\newblock Testing and modeling dependencies between a network and nodal attributes.
\newblock \emph{Journal of the American Statistical Association}, 110\penalty0 (511):\penalty0 1047--1056, 2015.

\bibitem[Frachetti et~al.(2017)]{frachetti2017nomadic}
M.~D. Frachetti et~al.
\newblock Nomadic ecology shaped the highland geography of {A}sia’s {S}ilk {R}oads.
\newblock \emph{Nature}, 543\penalty0 (7644):\penalty0 193--198, 2017.

\bibitem[Friedman et~al.(2008)Friedman, Hastie, and Tibshirani]{friedman2008sparse}
J.~Friedman, T.~Hastie, and R.~Tibshirani.
\newblock Sparse inverse covariance estimation with the graphical lasso.
\newblock \emph{Biostatistics}, 9\penalty0 (3):\penalty0 432--441, 2008.

\bibitem[Gelfand and Smith(1990)]{gelfand1990sampling}
A.~E. Gelfand and A.~F. Smith.
\newblock Sampling-based approaches to calculating marginal densities.
\newblock \emph{Journal of the American Statistical Association}, 85\penalty0 (410):\penalty0 398--409, 1990.

\bibitem[Gneiting et~al.(2007)Gneiting, Balabdaoui, and Raftery]{gneiting2007probabilistic}
T.~Gneiting, F.~Balabdaoui, and A.~E. Raftery.
\newblock Probabilistic forecasts, calibration and sharpness.
\newblock \emph{Journal of the Royal Statistical Society Series B: Statistical Methodology}, 69\penalty0 (2):\penalty0 243--268, 2007.

\bibitem[Graham(2020)]{graham2020dyadic}
B.~S. Graham.
\newblock Dyadic regression.
\newblock \emph{The Econometric Analysis of Network Data}, 2020\penalty0 (1):\penalty0 23--40, 2020.

\bibitem[Gr{\"u}nwald and Van~Ommen(2017)]{grunwald2017inconsistency}
P.~Gr{\"u}nwald and T.~Van~Ommen.
\newblock Inconsistency of {B}ayesian inference for misspecified linear models, and a proposal for repairing it.
\newblock \emph{Bayesian Analysis}, 12\penalty0 (4):\penalty0 1069--1103, 2017.

\bibitem[Hanks and Hooten(2013)]{hanks2013circuit}
E.~M. Hanks and M.~B. Hooten.
\newblock Circuit theory and model-based inference for landscape connectivity.
\newblock \emph{Journal of the American Statistical Association}, 108\penalty0 (501):\penalty0 22--33, 2013.

\bibitem[Hanks et~al.(2015)]{hanks2015restricted}
E.~M. Hanks et~al.
\newblock Restricted spatial regression in practice: {G}eostatistical models, confounding, and robustness under model misspecification.
\newblock \emph{Environmetrics}, 26\penalty0 (4):\penalty0 243--254, 2015.

\bibitem[Hefley et~al.(2017)]{hefley2017dynamic}
T.~J. Hefley et~al.
\newblock Dynamic spatio-temporal models for spatial data.
\newblock \emph{Spatial Statistics}, 20:\penalty0 206--220, 2017.

\bibitem[Hodges and Reich(2010)]{hodges2010adding}
J.~S. Hodges and B.~J. Reich.
\newblock Adding spatially-correlated errors can mess up the fixed effect you love.
\newblock \emph{The American Statistician}, 64\penalty0 (4):\penalty0 325--334, 2010.

\bibitem[Holmes and Walker(2017)]{holmes2017assigning}
C.~C. Holmes and S.~G. Walker.
\newblock Assigning a value to a power likelihood in a general {B}ayesian model.
\newblock \emph{Biometrika}, 104\penalty0 (2):\penalty0 497--503, 2017.

\bibitem[Hooten et~al.(2017)]{hooten2017animal}
M.~B. Hooten et~al.
\newblock \emph{Animal Movement: Statistical Models for Telemetry Data}.
\newblock CRC Press, 2017.

\bibitem[Hubisz et~al.(2020)Hubisz, Williams, and Siepel]{hubisz2020mapping}
M.~J. Hubisz, A.~L. Williams, and A.~Siepel.
\newblock Mapping gene flow between ancient hominins through demography-aware inference of the ancestral recombination graph.
\newblock \emph{PLoS Genetics}, 16\penalty0 (8):\penalty0 e1008895, 2020.

\bibitem[Hughes and Haran(2013)]{hughes2013dimension}
J.~Hughes and M.~Haran.
\newblock Dimension reduction and alleviation of confounding for spatial generalized linear mixed models.
\newblock \emph{Journal of the Royal Statistical Society Series B: Statistical Methodology}, 75\penalty0 (1):\penalty0 139--159, 2013.

\bibitem[Jombart(2008)]{jombart2008adegenet}
T.~Jombart.
\newblock adegenet: a {R} package for the multivariate analysis of genetic markers.
\newblock \emph{Bioinformatics}, 24\penalty0 (11):\penalty0 1403--1405, 2008.

\bibitem[Kelso and Patterson(2010)]{kelso2010introducing}
N.~V. Kelso and T.~Patterson.
\newblock Introducing natural earth data - naturalearthdata.com.
\newblock \emph{Geographia Technica}, 5\penalty0 (82-89):\penalty0 25, 2010.

\bibitem[Kenny et~al.(2020)Kenny, Kashy, and Cook]{kenny2020dyadic}
D.~A. Kenny, D.~A. Kashy, and W.~L. Cook.
\newblock \emph{Dyadic Data Analysis}.
\newblock Guilford Publications, 2020.

\bibitem[Kolaczyk and Cs{\'a}rdi(2014)]{kolaczyk2014statistical}
E.~D. Kolaczyk and G.~Cs{\'a}rdi.
\newblock \emph{Statistical Analysis of Network Data with R}, volume~65.
\newblock Springer, 2014.

\bibitem[Lindgren(2010)]{lindgren2010dyadic}
K.-O. Lindgren.
\newblock Dyadic regression in the presence of heteroscedasticity—an assessment of alternative approaches.
\newblock \emph{Social Networks}, 32\penalty0 (4):\penalty0 279--289, 2010.

\bibitem[Lu et~al.(2020)]{lu2020nonlinear}
X.~Lu et~al.
\newblock Nonlinear reaction--diffusion process models improve inference for population dynamics.
\newblock \emph{Environmetrics}, 31\penalty0 (3):\penalty0 e2604, 2020.

\bibitem[Majumder et~al.(2021)]{majumder2021statistical}
S.~Majumder et~al.
\newblock Statistical inference based on a new weighted likelihood approach.
\newblock \emph{Metrika}, 84:\penalty0 97--120, 2021.

\bibitem[Mallick et~al.(2023)]{mallick2023allen}
S.~Mallick et~al.
\newblock The {A}llen {A}ncient {DNA} {R}esource ({AADR}): {A} curated compendium of ancient human genomes.
\newblock \emph{bioRxiv}, pages 2023--04, 2023.

\bibitem[Matheson and Winkler(1976)]{matheson1976scoring}
J.~E. Matheson and R.~L. Winkler.
\newblock Scoring rules for continuous probability distributions.
\newblock \emph{Management Science}, 22\penalty0 (10):\penalty0 1087--1096, 1976.

\bibitem[McRae et~al.(2008)]{mcrae2008using}
B.~H. McRae et~al.
\newblock Using circuit theory to model connectivity in ecology, evolution, and conservation.
\newblock \emph{Ecology}, 89\penalty0 (10):\penalty0 2712--2724, 2008.

\bibitem[Miller(2004)]{miller2004tobler}
H.~J. Miller.
\newblock Tobler's first law and spatial analysis.
\newblock \emph{Annals of the Association of American Geographers}, 94\penalty0 (2):\penalty0 284--289, 2004.

\bibitem[Miller and Dunson(2019)]{miller2018robust}
J.~W. Miller and D.~B. Dunson.
\newblock Robust {B}ayesian inference via coarsening.
\newblock \emph{Journal of the American Statistical Association}, 114\penalty0 (527):\penalty0 1113--1125, 2019.

\bibitem[Narasimhan et~al.(2019)]{narasimhan2019formation}
V.~M. Narasimhan et~al.
\newblock The formation of human populations in {S}outh and {C}entral {A}sia.
\newblock \emph{Science}, 365\penalty0 (6457):\penalty0 eaat7487, 2019.

\bibitem[Nguyen et~al.(2019)]{nguyen2019comprehensive}
H.~Nguyen et~al.
\newblock A comprehensive survey of tools and software for active subnetwork identification.
\newblock \emph{Frontiers in Genetics}, 10:\penalty0 155, 2019.

\bibitem[Novembre et~al.(2008)]{novembre2008genes}
J.~Novembre et~al.
\newblock Genes mirror geography within {E}urope.
\newblock \emph{Nature}, 456\penalty0 (7218):\penalty0 98--101, 2008.

\bibitem[Patterson et~al.(2006)Patterson, Price, and Reich]{patterson2006population}
N.~Patterson, A.~L. Price, and D.~Reich.
\newblock Population structure and eigenanalysis.
\newblock \emph{PLoS Genetics}, 2\penalty0 (12):\penalty0 e190, 2006.

\bibitem[Patterson et~al.(2012)]{patterson2012ancient}
N.~Patterson et~al.
\newblock Ancient admixture in human history.
\newblock \emph{Genetics}, 192\penalty0 (3):\penalty0 1065--1093, 2012.

\bibitem[Petkova et~al.(2016)Petkova, Novembre, and Stephens]{petkova2016visualizing}
D.~Petkova, J.~Novembre, and M.~Stephens.
\newblock Visualizing spatial population structure with estimated effective migration surfaces.
\newblock \emph{Nature Genetics}, 48\penalty0 (1):\penalty0 94--100, 2016.

\bibitem[Puu(1981)]{puu1981structural}
T.~Puu.
\newblock Structural stability and change in geographical space.
\newblock \emph{Environment and Planning A}, 13\penalty0 (8):\penalty0 979--989, 1981.

\bibitem[Roberts and Rosenthal(2009)]{roberts2009examples}
G.~O. Roberts and J.~S. Rosenthal.
\newblock Examples of adaptive {MCMC}.
\newblock \emph{Journal of Computational and Graphical Statistics}, 18\penalty0 (2):\penalty0 349--367, 2009.

\bibitem[Rohland et~al.(2022)]{rohland2022three}
N.~Rohland et~al.
\newblock Three assays for in-solution enrichment of ancient human {DNA} at more than a million {SNP}s.
\newblock \emph{Genome Research}, 32\penalty0 (11-12):\penalty0 2068--2078, 2022.

\bibitem[Rue and Held(2005)]{rue2005gaussian}
H.~Rue and L.~Held.
\newblock \emph{Gaussian Markov Random Fields: Theory and Applications}.
\newblock CRC Press, 2005.

\bibitem[Schmid and Schiffels(2023)]{schmid2023estimating}
C.~Schmid and S.~Schiffels.
\newblock Estimating human mobility in {H}olocene {W}estern {E}urasia with large-scale ancient genomic data.
\newblock \emph{Proceedings of the National Academy of Sciences}, 120\penalty0 (9):\penalty0 e2218375120, 2023.

\bibitem[Schwob et~al.(2019)Schwob, Zhan, and Dempsey]{schwob2019modeling}
M.~R. Schwob, J.~Zhan, and A.~Dempsey.
\newblock Modeling cell communication with time-dependent signaling hypergraphs.
\newblock \emph{IEEE/ACM Transactions on Computational Biology and Bioinformatics}, 18\penalty0 (3):\penalty0 1151--1163, 2019.

\bibitem[Schwob et~al.(2023)Schwob, Hooten, and McDevitt-Galles]{schwob2023dynamic}
M.~R. Schwob, M.~B. Hooten, and T.~McDevitt-Galles.
\newblock Dynamic population models with temporal preferential sampling to infer phenology.
\newblock \emph{Journal of Agricultural, Biological and Environmental Statistics}, 28\penalty0 (4):\penalty0 774--791, 2023.

\bibitem[Simon et~al.(2013)]{simon2013sparse}
N.~Simon et~al.
\newblock A sparse-group lasso.
\newblock \emph{Journal of Computational and Graphical Statistics}, 22\penalty0 (2):\penalty0 231--245, 2013.

\bibitem[Teller(1937)]{teller1937crossing}
E.~Teller.
\newblock The crossing of potential surfaces.
\newblock \emph{Journal of Physical Chemistry}, 41\penalty0 (1):\penalty0 109--116, 1937.

\bibitem[Wang(2013)]{wang2013examining}
I.~J. Wang.
\newblock Examining the full effects of landscape heterogeneity on spatial genetic variation: a multiple matrix regression approach for quantifying geographic and ecological isolation.
\newblock \emph{Evolution}, 67\penalty0 (12):\penalty0 3403--3411, 2013.

\bibitem[Warren et~al.(2023)]{warren2023spatial}
J.~L. Warren et~al.
\newblock Spatial modeling of \textit{{M}. tuberculosis} transmission with dyadic genetic relatedness data.
\newblock \emph{Biometrics - Early Preview}, 2023.

\bibitem[Wessel et~al.(2019)]{wessel2019generic}
P.~Wessel et~al.
\newblock The {G}eneric {M}apping {T}ools version 6.
\newblock \emph{Geochemistry, Geophysics, Geosystems}, 20\penalty0 (11):\penalty0 5556--5564, 2019.

\bibitem[Wikle and Hooten(2010)]{wikle2010general}
C.~K. Wikle and M.~B. Hooten.
\newblock A general science-based framework for dynamical spatio-temporal models.
\newblock \emph{Test}, 19:\penalty0 417--451, 2010.

\bibitem[Wikle et~al.(2001)]{wikle2001spatiotemporal}
C.~K. Wikle et~al.
\newblock Spatiotemporal hierarchical {B}ayesian modeling tropical ocean surface winds.
\newblock \emph{Journal of the American Statistical Association}, 96\penalty0 (454):\penalty0 382--397, 2001.

\end{thebibliography}

\end{document}